\documentclass[10pt]{article}
\usepackage{graphics}

\usepackage{graphicx}
\input{psfig.sty}

\usepackage{amstext,amssymb,amsmath}
\usepackage{epsfig,dcolumn}
\usepackage{graphicx}
\usepackage{graphicx}
\usepackage{bm}
\usepackage{epsfig,dcolumn}
\usepackage[usenames]{color}

\def\lsim{\mathrel{\rlap{\lower4pt\hbox{\hskip1pt$\sim$}}
    \raise1pt\hbox{$<$}}}
\def\gsim{\mathrel{\rlap{\lower4pt\hbox{\hskip1pt$\sim$}}
    \raise1pt\hbox{$>$}}}

\begin{document}

\noindent
\begin{center}
\ \center\small{COULOMB GAUGE  QCD AND THE EXCITED HADRON SPECTRUM}
\ \center\small{FELIPE LLANES-ESTRADA$^a$, STEPHEN R. COTANCH $^b$, TIM VAN CAUTEREN$^a$, JUAN M. TORRES-RINCON$^a$,  PEDRO BICUDO$^c$ and MARCO CARDOSO$^c$}
\ \center{$^a$\em{Depto. F\'isica Te\'orica I, Universidad Complutense de Madrid,
28040 Madrid, Spain}}
\ \center{$^b$\em{Department of Physics, North Carolina State University, Raleigh, 
NC 27695, USA}}
\ \center{$^c$\em{Instituto Superior Tecnico, Avda. Rovisco Pais 1096, Lisbon, Portugal}}
\end{center}


\noindent
We discuss  progress in understanding the light and heavy quark excited hadron spectrum from  Coulomb
gauge  QCD.  For light quark  systems we highlight  the insensitivity to spontaneous chiral symmetry breaking, which predicts Wigner parity-degeneracy in the highly excited hadron spectrum and  allows the  quark mass momentum dependence to be experimentally probed.
For heavy quark meson decays we invoke the Franck-Condon principle, a consequence of small velocity changes for heavy quarks, to extract qualitative, but model independent,   structure
insight from the momentum distribution of the decay products.
 \\
\hspace{-1.5em}PAC numbers: 11.30.Rd, 12.38.Lg, 12.39.Mk, 12.40.Vv, 12.40.Yx
\\
\
\hspace{-1.5em}Keywords:  Coulomb gauge QCD,  excited hadrons, chiral symmetry
\\

\begin{center}
\center{\Large\em{1.  Motivation for Coulomb gauge QCD  studies }}
\\
\end{center}


Recent years have witnessed steady progress in both understanding and applying Quantum Chromodynamics in Coulomb gauge. Much of the attention has been formal, focusing on the  such issues  as the behavior of the Green's function in the deep infrared~\cite{Tubingennew}, the Slavnov-Taylor identities~\cite{WatsonReinhardt} necessary for a complete renormalization treatment  in this gauge and the structure of the ground state vacuum~\cite{Adamnew}.
In this paper the thrust is more phenomenological as we detail how an approximate
Coulomb gauge formulation~\cite{flsc} can provide important hadron structure insight.
 While the QCD treatment  is not rigorous, we have been able to identify several fundamental issues which a more precise  Coulomb gauge QCD analysis can resolve.

In addition to the advantages of Coulomb gauge QCD, it is worthwhile to also comment on this formulation's major limitation  involving calculating observables that connect  different reference frames.  Since the Coulomb condition, $\nabla\cdot {\bf A}^a=0$, is specified for one chosen frame (typically the hadron center of momentum),  the computation of a form factor or an electromagnetic transition matrix element, which requires hadron wavefunctions in two different frames, entails employing  boost operators~\cite{rocha}. This introduces a formidable difficulty since in this gauge these operators are as complicated as the exact Hamiltonian. Although one can adapt methods developed for the relativistic quark model~\cite{cotanchboosts}, unambiguous, reliable results become very difficult to obtain. Related,  fragmentation functions and parton distribution functions, more  conveniently calculated in light front quantization, are also difficult to formulate in the Coulomb gauge.

In contrast,  Coulomb gauge QCD is much more ideal for spectroscopy and hadron structure including the vacuum which is difficult for light cone formulations.
Solving the equation of motions variationally either on the lattice or by approximate model truncation allows study of the entire spectrum with masses of arbitrary excitations (in the rest frame), any spin and all other quantum numbers.
The variational treatment is possible because this gauge is physical (contains no spurious degrees of freedom or ghosts),
involving only transverse gluons.  Gauss's law, which is essential for confinement,  is 
explicitly satisfied and incorporated as  a constraint. However, the resulting  interaction kernel is complicated. Fortunately,  with  insight provided by lattice QCD, it is amendable to modeling using a more calculable confining kernel containing a linear potential as used in
Ref. \cite{flsc} and further supported by other approaches \cite{linearpotential}.
This approximate framework provides several new insights into the structure of highly excited mesons and baryons. One is the role of chiral symmetry for the complete
 hadron spectrum. As detailed in section 2, because chiral symmetry is incorporated (unlike 
 constituent quark model formulations), one can clearly understand why the light quark, low mass pseudoscalar mesons  are strongly governed by spontaneous chiral symmetry breaking while the high lying spectrum is insensitive to chiral symmetry breaking leading to parity doublets.
   This is difficult to
 explain with  other methods which generally  do not respect chiral symmetry and/or 
can not predict the excited spectrum. 
For example, for heavy quarks  non-relativistic QCD (NRQCD)~\cite{vairo} reasonably describes the low lying spectrum. However, it becomes inadequate for strongly excited states due to the 
attending large quark momenta. Again, the Coulomb gauge formulation naturally extends to systems involving arbitrary large momenta. 
Further, in addition to mass spectra, one can also learn about the momentum structure of the hadrons in strong decays, due to the large scale provided by the heavy quark mass and high momentum, $m_Q v\gg \Lambda_{\rm QCD}$, which is discussed in section 3. This  provides decay signatures which may be useful for identifying exotic hadrons and  is analogous to the Franck-Condon principle of molecular physics. Section 4 discusses new insights through future measurements at Jefferson Lab with the anticipated 12 GeV
upgrade. Finally a summary of key results provided in this paper is presented in section 5.
\\


\begin{center}
\center{\Large\em{2. Insight into chiral symmetry   }}
\\
\end{center}


Chiral symmetry is said to be hidden (or spontaneously broken) by the condensation of quarks in the ground state (vacuum) of Quantum Chromodynamics. This feature, documented in both Landau and Coulomb gauge studies, is expected to be of lesser importance in the excited spectrum as 
illustrated by a simple analogy.
Consider two identical solid bridge supporting pillars in two different shallow  ponds, one salt and  the other fresh water (thus breaking the symmetry between them). Their net weight is essentially the same (approximately symmetrical) even though one experiences a slightly larger buoyant force from the salt water. Two highly excited, but opposite parity, hadrons are analogous to the two pillars  since their quarks  have (on average) large momenta but now the running quark mass is small \cite{VanCauteren:2009vm}, so chiral symmetry is approximately restored.
The first statement, that the momentum is larger for excited states, is supported by the relativistic virial theorem~\cite{Lucha:1990mm}
and also explicit variational model computations for momentum distributions as detailed in our Coulomb gauge approach~\cite{Bicudo, Bicudo:2009cr}.
The second statement, that the running quark mass is small for large momentum, is a consequence of asymptotic freedom, well-known from early QCD studies~\cite{Adler:1984ri}.

\begin{figure}[!b]
\vspace{-4cm}
\centering
\includegraphics[scale=.33]{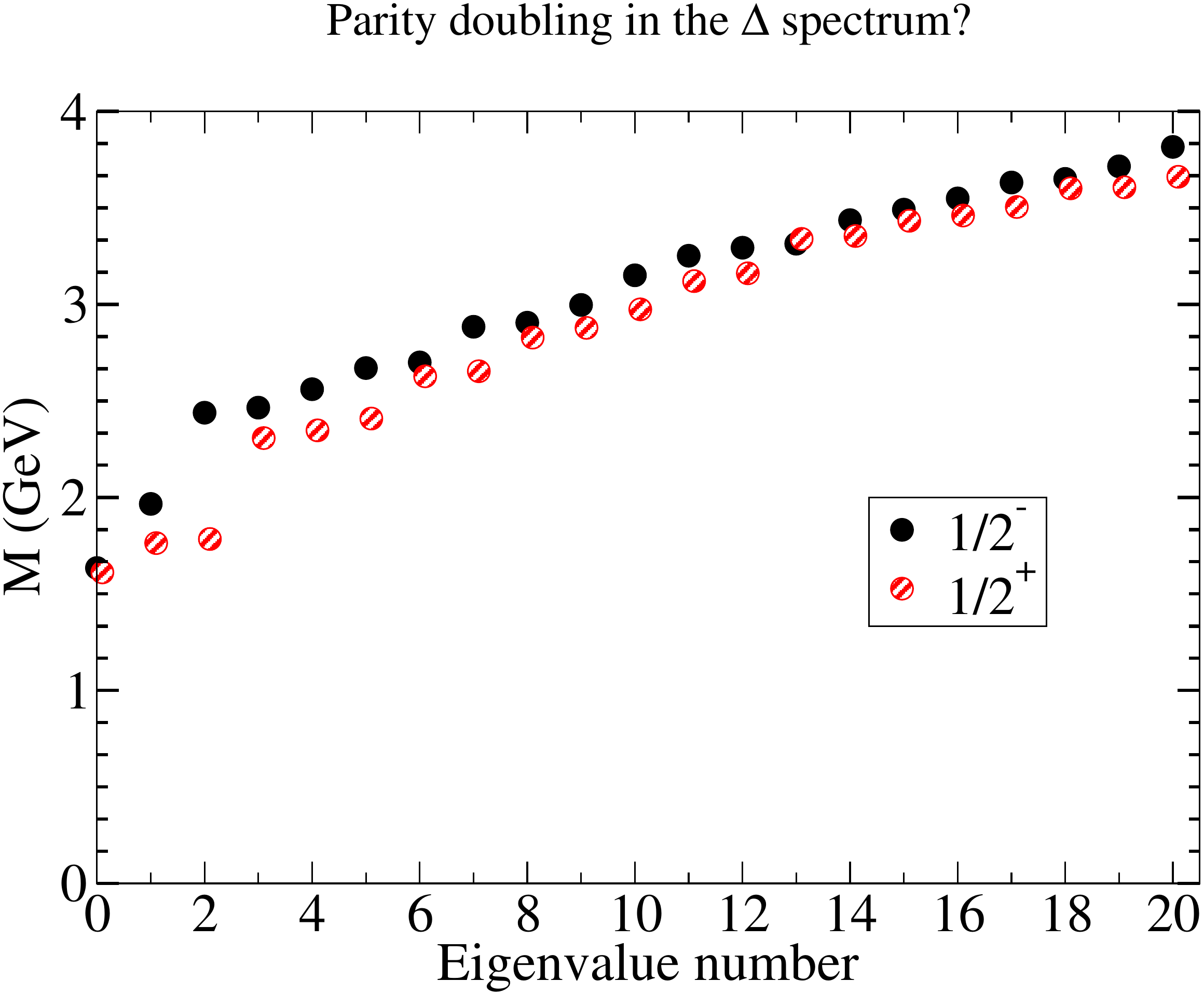}
\caption{
Excited $I=3/2$, $J=1/2$ $\Delta$ spectrum.  Parity doublets emerge, especially for higher excited states.}
\label{fig:spectrum}
\end{figure}

An immediate consequence of chiral symmetry restoration is the reappearance of Wigner symmetry in the highly excited spectrum, with doublet partner states of equal mass but opposite parity \cite{Glozman:1999tk}. 
For three quark configurations, relativistic chiral approaches
like ours predict that states  come in
quartets, reducible representations that split into two doublets [15].
As an example, 
take the family of spin $1/2$ excitations of the isospin $3/2$ $\Delta$.
We have used the Cornell approach to Coulomb gauge QCD with harmonic oscillator interactions and a large, harmonic oscillator basis model space  to diagonalize the Hamiltonian.
All orthogonalizations and 
symmetrizations are performed numerically and the resulting $\Delta$ spectrum is plotted in Fig.~\ref{fig:spectrum} for the first 20 eigenvalues.
Note that, while there are  variations for the lower states, parity doubling is indeed a feature of the higher spectrum illustrating  chiral symmetry in the Wigner mode. Lattice gauge simulations and also Dyson-Schwinger approaches have difficulties generating excited state spectra which is easily computed in Coulomb gauge based models.

\newpage

\begin{figure}[!h]
\centering
\includegraphics[width=0.75\columnwidth]{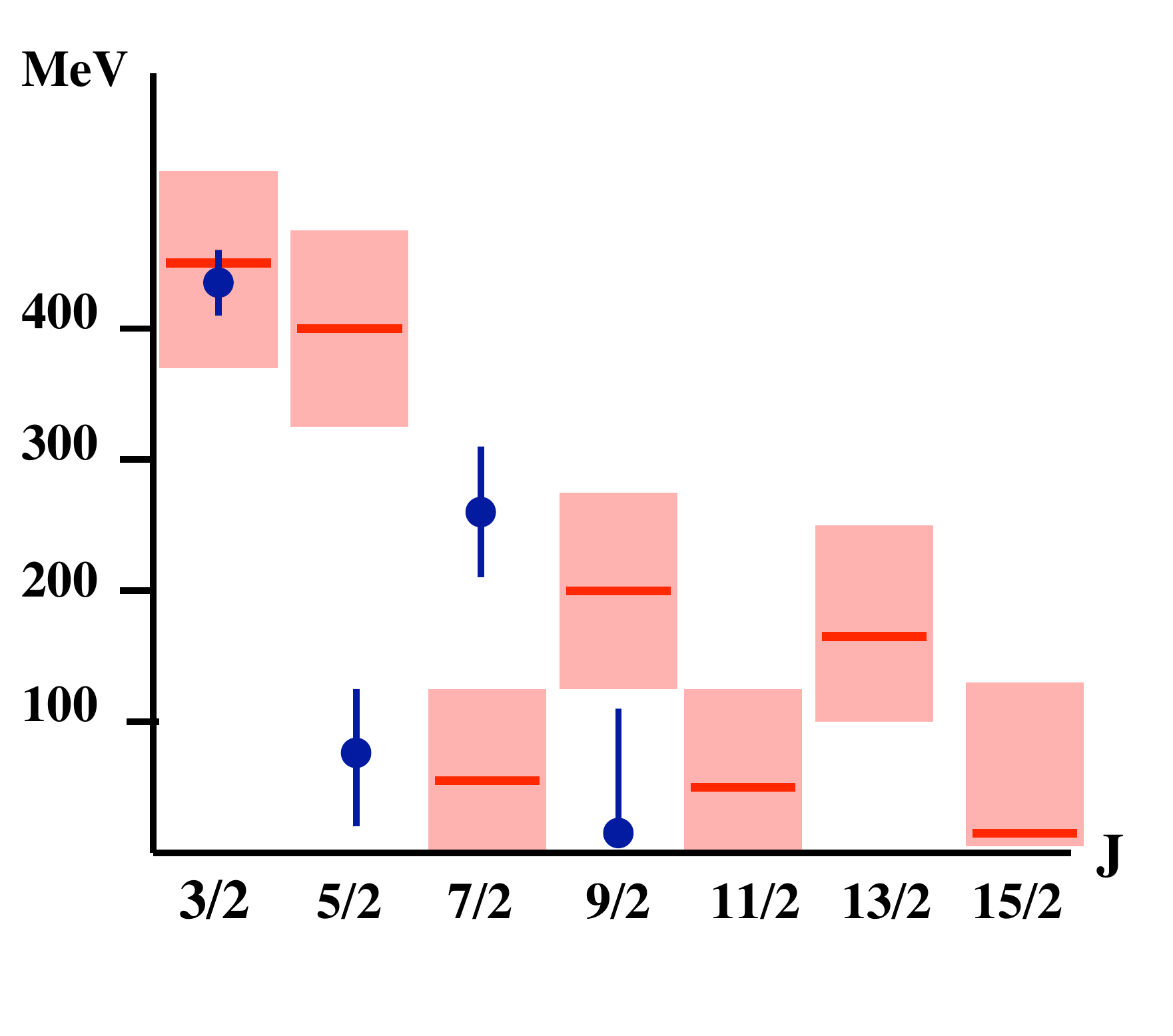}
\caption{
Predicted baryon spectrum from variational Monte Carlo  calculations (boxes) compared to  data (solid circles) from  Ref.~\cite{pdg}.}
\label{fig:measureM1}
\end{figure}

A potential problem investigating parity doublets experimentally, especially if the spectrum is numerous,
is to identify the correct corresponding Wigner partners. This is where theoretical calculations, which do not
have this ambiguity, can be of great benefit as we again document in  Fig. \ref{fig:measureM1}   where the doublet separation energy (splitting) is plotted for each $J$.
The parity degeneracy clearly emerges for increasing excitation energy and $J$. 
This variational Monte Carlo investigation is still in progress and the attending error bars 
have not been shown.

Related, but perhaps more interesting,  one can use the onset of this degeneracy to probe the quark mass momentum dependence (running) in the infrared.
The quark mass in our approach appears both in the quark spinors and in
the QCD Hamiltonian. For small quark mass
and high momentum, a series expansion of the spinors and the Hamiltonian in the 
parameter $m(k)/k$ reveals several insights. 
Expanding the spinors
 \begin{equation}
 u_{\lambda}({\bf k}) = \frac{1}{2E(k)} \left[ \begin{array}{c} \sqrt{E(k) +
m(k)} \chi_\lambda \\ \sqrt{E(k) - m(k)} \, {\mbox{\boldmath$\sigma$\unboldmath}}\cdot \hat{{\bf k}} \chi_\lambda  \end{array} \right]  \ ,
\end{equation}  
yields  the leading terms at high momentum
\begin{equation}
u_{\lambda}({\bf k})  \rightarrow  \frac{1}{\sqrt{2}} \left[ \begin{array}{c}
\chi_\lambda \\ \, {\mbox{\boldmath$\sigma$\unboldmath}} \cdot \hat{{\bf k}} \chi_\lambda \end{array} \right] +
\frac{1}{2\sqrt{2}} \frac{m(k)}{k} \left[ \begin{array}{c} \chi_\lambda \\
-\, {\mbox{\boldmath$\sigma$\unboldmath}} \cdot \hat{{\bf k}} \chi_\lambda \end{array} \right]  \, , 
\end{equation}
with $E(k) = \sqrt{k^2 + m(k)^2}$. The first term is chirally invariant
and the second chiral-symmetry breaking. Note that the lower component  in each term has
the opposite sign.
Instead of directly expanding the QCD Hamiltonian, expand the Hamiltonian matrix elements computed in the Hilbert space spanning
highly excited resonances, labeled $n$, where the average momentum, $<k>$, is large
\begin{equation}
 \langle n | H^{QCD} | n \rangle = \langle n |
H^{QCD}_{{\chi}_S }| n \rangle + \langle n | \frac{m(k)}{k} H^{QCD}_{\chi_B} |
n \rangle + \dots  \label{QCDexp} \; .
\end{equation}
 Again the first  Hamiltonian term, labeled by $\chi_S$, is chirally symmetric while the second,
 labeled $\chi_B$, is not and  
involves nonchiral, spin-dependent potentials in the quark-quark interaction.

Now recall for zero mass current quarks, the chiral charge,
\begin{equation}
 Q_5 = \int d{\bf x} \,  \psi^\dagger({\bf x}) \gamma_5 
\psi({\bf x}) \ ,
\end{equation}
 commutes with the QCD Hamiltonian. Nevertheless, chiral symmetry is
still  spontaneously broken by the ground state vacuum since $Q_5 | 0 \rangle \neq 0$,
leading to a large constituent quark mass in the quark propagator, pseudoscalar Goldstone bosons
and no parity-degeneracy in ground-state baryons. Insight about the chiral charge operator follows by performing a BCS transformation  to a quasiparticle basis with operators $B$ and $D$ 
defined by
\begin{eqnarray} \label{eq:operator rotations}
    B_{\lambda {\cal C}}({\bf k}) &=& \cos\frac{\theta_k}{2}
    b_{\lambda {\cal C}}({\bf k}) - \lambda \sin\frac{\theta_k}{2}
    d^\dag_{\lambda {\cal C}}({\bf -k})    \\ 
    D_{\lambda {\cal C}}({\bf -k})&=& \cos\frac{\theta_k}{2}
    d_{\lambda {\cal C}}({\bf -k})  + \lambda \sin\frac{\theta_k}{2}
    b^\dag_{\lambda {\cal C}}({\bf k})   \ .
   \end{eqnarray}
Then using the standard Fock operator normal mode representation for the quark field
\begin{equation}
\label{colorfields1}
 \Psi({\bf x}) =\sum_{\lambda {\cal C}}\int \!\! \frac{d
    {\bf k}}{(2\pi)^3} [{u}_{\lambda} ({\bf k}) b_{\lambda {\cal C}}({\bf k})  + {v}_{\lambda} (-{\bf k})
    d^\dag_{\lambda {\cal C}}({\bf -k})] e^{i {\bf k} \cdot {\bf x}} \hat{{\epsilon}}_{\cal C} \ ,
    \end{equation}
and substituting for the bare spinors, the chiral charge can be expressed as
\begin{eqnarray} \label{chiralcharge}
Q_5  &=& \int \frac{d \bf k}{(2\pi)^3} \sum_{\lambda
\lambda '{\cal C}} 
{ k \over \sqrt{ k^2 + m^2(k)}}
 \\ \nonumber 
&&\times \left[ (\, {\mbox{\boldmath$\sigma$\unboldmath}}\cdot{\bf
\hat{k}})_{\lambda \lambda'}  	
\left( B^\dagger_{ \lambda {\cal C}} ({\bf k})B_{ \lambda' {\cal C}}({\bf k}) + D^\dagger_{ \lambda' {\cal C}}({\bf -k})
D_{ \lambda{\cal C}}({\bf -k})
\right) + \right. \\  \nonumber 
&&\left.
{ m(k) \over k} (i\sigma_2)_{\lambda \lambda'} \
\left( B^{\dagger}_{\lambda {\cal C}}({\bf k}) D^\dagger_{\lambda' {\cal C}}({\bf- k}) +
B_{ \lambda' {\cal C}}({\bf k}) D_{ \lambda {\cal C}}({\bf -k})
\right) \right] \ .
\end{eqnarray}
The first two terms between the square brackets are quark and anti-quark 
number operators, i.e. operators conserving the number of particles,  flipping spin and parity. For $m(k) << k$, it dominates the
third and fourth terms representing pion creation and annihilation. 
As  argued in Ref.~\cite{Bicudo:2009cr},
repeated action of the chiral
charge on a three-quark state produces a quartet of states, two of each parity,
that dynamically split into two doublets of parity partners. Moreover, the mass splitting between
partners is a direct measure of $m(k)$ and they
are degenerate when $m(k)$ vanishes.

\begin{figure}[t]
\centering
\includegraphics[scale=.45]{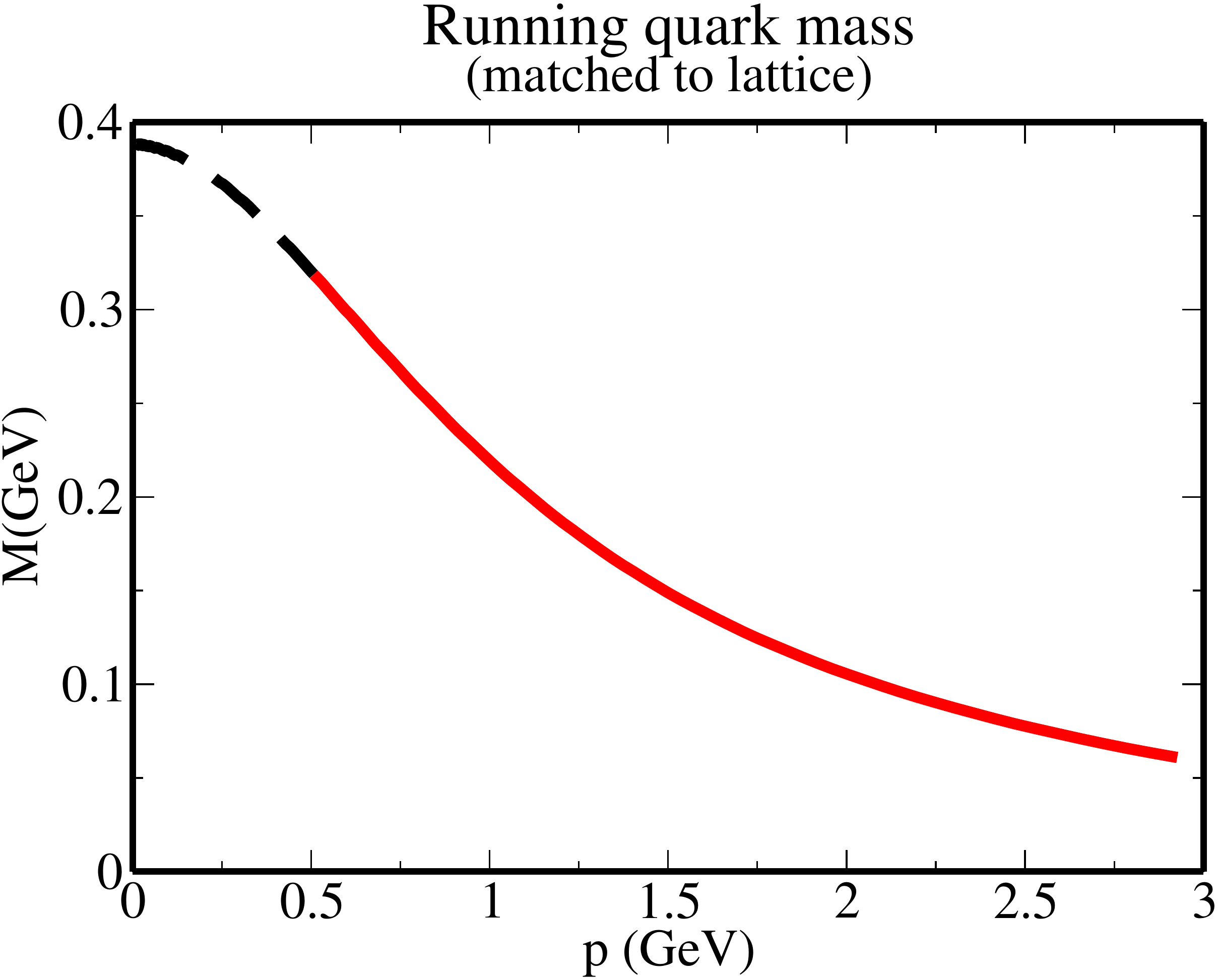}
\caption{
Cartoon showing the  running quark mass as a function of momentum.
The solid line is a power law whose exponent can be
obtained from the decreasing parity doublet  splittings for increasing
$J$, as illustrated in Fig. ~\ref{fig:measureM1}.}
\label{fig:measureM2}
\end{figure}

In order to link this parity doublet mass splitting, $\Delta M =|M^{P=+} - M^{P=-}|$,  to
the running quark mass, we examine  the lowest-lying  parity
doublets for increasing spin $J$ and incorporate the following key elements.
\begin{itemize}
 \item Regge trajectories, $J = \alpha_0 + \alpha {M^{\pm}}^2
    \, \stackrel{_{J\to\infty}}{\longrightarrow} \, \alpha {M^{\pm}}^2$.
 \item Relativistic virial theorem, $\langle k \rangle \to c_2 M^\pm
\to \frac{c_2}{\sqrt{\alpha}} \sqrt{J}$.
 \item Canceling of the chirally invariant term, $<n|H_{\chi S}^{QCD}|n>$,  in $\Delta
M$ so that $\Delta M << M^\pm $
 and
      $\Delta
M \to \,
      <\frac{m(k)}{k} {H_{\chi B}^{QCD}}> \, \to \, c_3 \frac{m(<k>)}{<k>}
      <{H_{\chi B}^{QCD}}>$.
 \item In $H_{\chi S}^{QCD}$, the spin-orbit term, ${\bf L} \cdot {\bf S}$,
    is crucial to correct the angular momentum in the centrifugal
    barrier  from ${\bf L}^2$ to the chirally invariant
    ${\bf L}^2 + 2{\bf L}_ \cdot {\bf S} = {\bf J}^2 -
    \frac{3}{4}$. Due to the sign difference in the
    helicity-dependent term, $ - {\bf \sigma}\cdot{\bf
\hat{k}}$, in the spinor, the spin-orbit term in
    $H^{QCD}_{\chi B}$ adds to the mass splitting  instead
    of cancelling out as it does for $H^{QCD}_{\chi S}$. Since the centrifugal
barrier scales like $M^\pm$ for high $J$, the
    spin-orbit term scales with one power of $J$ less
    \begin{equation}
     <H^{QCD}_{\chi B}> \to c_5 M^\pm J^{-1} \to
    \frac{c_5}{\sqrt{\alpha}} \sqrt{\frac{1}{J}} \ . \label{eq:tensor} \nonumber
\end{equation}
\end{itemize}
Combining these elements yields 
\begin{equation} 
 \Delta M \to \frac{c_3 c_5}{c_2
    \sqrt{\alpha}} m(<k>) J^{-1} \; .
\end{equation}
This equation  linearly relates the parity splitting (an observable) to the running of the quark mass at the average momentum $<k>$ for that splitting.  Note that $<k>$ will vary with splitting, increasing for higher excited states (and $J$).
Although there are still unknown constants that are gauge dependent, a useful  power-law running of $m(<k>)$ (or equivalently $m(k)$) can be obtained model independently from  data at high spin $J$
 by fitting 
the splitting  to the  form $\Delta M \propto J^{-i}$ and determining the exponent $i$.
Then, the power-law behaviour of the
running quark mass is  given by 
\begin{equation} \label{key2}
m(k) \propto k^{-2i+2}  \ . 
\end{equation}
As indicated in Fig. \ref{fig:measureM1},  the
known masses of lowest-lying $\Delta$ resonances for $J=1/2,\cdots,15/2$  is currently not
sufficient to derive this exponent. Establishing the masses of the parity
doublets for spins $J >9/2$ would resolve this and provide a running quark mass
momentum dependence as schematically indicated in Fig.~\ref{fig:measureM2}.

\begin{center}
\center{\Large\em{3. Franck-Condon principle and heavy hadron decays }}
\\
\end{center}


In general, calculating most hadron observables
such as masses, widths, or partial branching ratios
 entails integrating over quark momenta which
precludes direct, experimental  study of momentum distributions.
Here, we submit that 
it is possible to access this  distribution by measuring the decay hadron momenta of 
open flavor, light-heavy quark $q \bar Q$ and $\bar q Q$ systems
from  highly excited heavy quarkonium $Q \bar Q$ states. This assertion is based upon two points.
The first is that $Q \bar Q$ states with high excitation energies also have high momenta.
 The second is the heavy quark  "velocity superselection rule" which  states
 that in  heavy quark meson decays the velocity of the 
heavy quark is conserved and  not affected by the non-heavy degrees of freedom (light quarks and gluons).
This selection rule is a common ingredient in  heavy quark effective theories such as HQET
and NRQCD and is due to the clear scale separation between
 the large mass of the heavy quark and $\Lambda_{QCD}$ (as well as the running masses of light quarks).  In particular,
 the NRQCD Lagrangian at leading order in $\Lambda_{QCD}/m_{Q}$ for  heavy   quarkonium is
\begin{equation}
 \mathcal{L}_{LO} =  
 \left( \begin{array}{c}  \overline{h}_v^{(+)} , \; \overline{h}_w^{(-)}   \end{array} \right)  
 \left( \begin{array}{cc} i {\bf v} \cdot {\mbox{\boldmath$\nabla$\unboldmath}} - \frac{\nabla^2}{2m_Q}  & 0 \\ 0 & i {\bf w} \cdot {\mbox{\boldmath$\nabla$\unboldmath}} - \frac{\nabla^2}{2m_Q}      \end{array} \right) \\ \left( \begin{array}{c} h_v^{(+)}  \\  h_w^{(-)}   \end{array} \right)  , 
 \end{equation}
where  the velocities, $v$ and $w$,  are 
the same for the quark creation and destruction operators, $h^{(\pm)}$ . It is clear that the operator structure of this Lagrangian preserves the heavy quark velocity.
Coulomb gauge QCD also takes this form in the heavy quark limit as can
be seen from the explicit Hamiltonian \cite{flsc}.

Returning to accessing quark momentum distributions and invoking these two principles for a highly excited $Q \bar Q$  decay  to open flavor mesons, $\rightarrow \bar q Q  q \bar Q $,
in leading order,  the velocity (and momentum) of each heavy quark  is the same before and after the decay. Then the quarkonium  momentum distribution is approximately
 given by the momentum of the final state  mesons containing the heavy quarks, which is directly measurable. This only applies to 3 or 4-body final states since  2-body final states 
 have fixed momentum due to over all momentum conservation. Technically,
in the limit of infinite heavy quark mass these principles, combined with momentum conservation, would forbid a two-body
final state decay. This implies that for finite mass, decays to two-body states should be suppressed
relative to three or more hadron final states.   The above decay features should therefore be more
pronounced in bottomonium transitions compared to charmonium decays.


This effect \cite{TorresRincon:2010fu} is very similar to the well-known Franck-Condon principle in molecular physics. It states that during an electronic 
transition, the nuclear degrees  freedom (position and momentum) are essentially frozen due to the large scale
 difference between the electron and nuclei masses. Therefore, the wavefuncion associated with  the motion of the nuclei remains the same after the transition.
 
 It is thus possible to  reconstruct the excited state quarkononium wavefunction
by measuring   the final state  heavy meson
momentum distribution.
In particular, a highly excited pure $Q\overline{Q}$ state has a definite structure of Sturm-Liouville zeroes (nodes) in the wave function, whereas a $Q {\bar Q} g$
 hybrid state (that is, quarkonium plus a gluonic excitation) with the same mass, say in the ground state, would have no Sturm-Liouville zeroes (no  nodes). Experiment can now  distinguish between these two hadron structures.
 
\begin{figure}
\begin{center}
\hspace{-1cm}
\includegraphics[scale=.55]{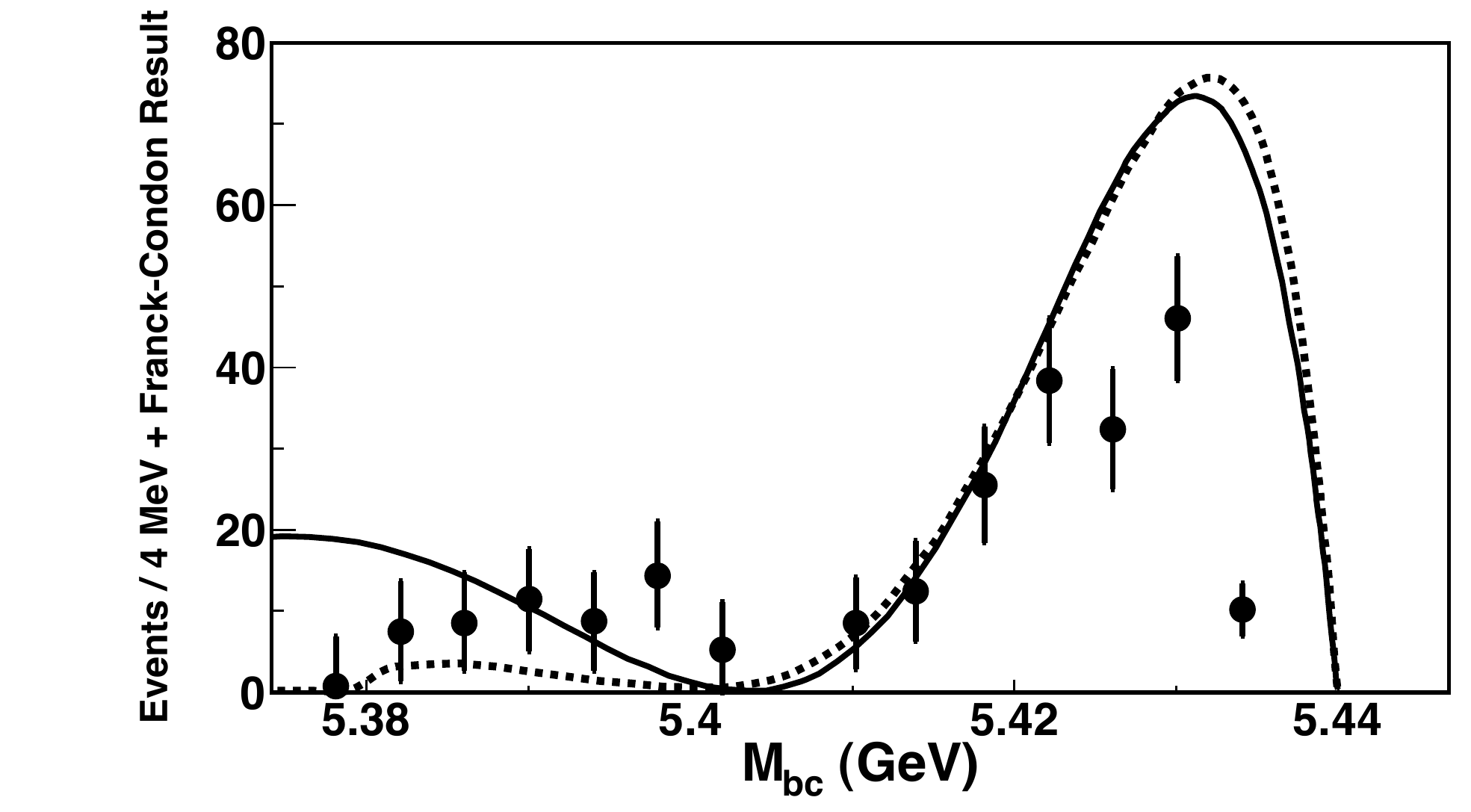}
  \caption{Decay widths for $\Upsilon(10860)$. Solid  line: predicted momentum distribution using the Frank-Condon principle for  a $5S$ quarkonium wavefunction
   multiplied by phase space factors. Dotted line is the same as solid but now a decay mechanism is included. The dip in the data (dots) \cite{Drutskoy:2010an} is correlated with the first node in the momentum distribution.
\label{figDrutskoy}
}
\end{center}
\end{figure}

 As an illustrative example consider the decay of $\Upsilon(10860)$  to
 $B_s$ $\overline{B}_s$ and $B^*_s$ $\overline{B}^*_s$ final states.  This parent state is
 generally accepted as an excited two-body quarkonium configuration, i.e. having several  nodes.
The momentum distribution for this state was calculated \cite{TorresRincon:2010fu}  using a Coulomb gauge model  $5S$ wavefunction having four nodes and is plotted in Fig.  \ref{figDrutskoy}. Also plotted (dotted line) is the same distribution but now
 a decay mechanism is included (${^3} P_0$ scalar model).
 Both distributions have also been multiplied by
 phase space factors and exhibit a clear node
 structure (depicting  additional nodes requires a non-linear graph). Also plotted is the recent
 Belle Collaboration data \cite{Drutskoy:2010an}  for $\Upsilon(10860)$
 decay to $B_s$ $\overline{B}_s$ and $B^*_s$ $\overline{B}^*_s$ including 2, 3 and 4-body final meson  states. Notice that there is reasonable agreement, especially the 
 correspondence for the first Sturm-Liouville zero,  affirming the validity of the
 Frank-Condon principle for heavy quark systems.

Finally, one can also construct a different momentum signature to distiguish between $Q \overline{Q} g$ hybrid  candidates and  possible tetraquark states with similar mass (both without Sturm-Liouville zeroes).
Now it is necessary to study  four-meson decays  such as $Z \rightarrow D \overline{D} D \overline{D}$, or $D \overline{K} K \overline{D}$ for which one can obtain, from the momentum distribution
 in the final state, the off-plane correlator, 
 
\begin{equation}
\Pi (\mathbf{p}_1,\mathbf{p}_2,\mathbf{p}_3, \mathbf{p}_4)=\frac{|(\mathbf{p}_1 \times \mathbf{p}_2)\cdot \mathbf{p}_3|^2}
{ {\displaystyle \prod_{\substack{
i, j=1 \\
i \neq j}}^4 \ [|\mathbf{p}_i \times \mathbf{p}_j|]^{1/2}} } \ ,
\end{equation}
that  reflects a distinctive  non-planar structure from  momenta combinations not present in 2-body or 3-body states.  Consult  Ref.  \cite{TorresRincon:2010fu} for a more complete discussion on using this signature to identify  tetraquark states.
\\

\begin{center}
\center{\Large\em{4. Future excited spectrum studies  at Jefferson Lab }}
\\
\end{center}


As discussed at this conference, one of the goals for the 12 GeV upgrade of CEBAF is to explore the highly excited meson and baryon spectrum with existing experimental halls also having upgraded instrumentation.  This will facilitate electroproduction measurements of excited baryon resonances with higher spins, $J$,  allowing completion of the analysis involving Eq.~\ref{key2}.
Further, finding these excited states will bring closure to quark models which
have long predicted their existence.  There is also a QCD motivation, to understand the onset of chiral symmetry breaking in the low spectrum, by clearly linking $m(k)$ from the asymptotically free regime to the constituent quark mass. This is possible because $m/k$ is a small parameter in the excited spectrum.

Equally, if not more,  important is the construction of the new Hall D to conduct Glue-X experiments
to search for exotic  hadrons with gluonic degrees of freedom.  This hall will also permit exploring the  excited meson spectrum, complimenting  the other  halls in this endeavor.
Since discovering exotica is a major goal of Glue-X, having a constraint on  momentum distributions would be very useful and this may be possible again using the Frank-Condon principle.
Consider for example the  $\phi(2170)$ observed in electron-positron collisions by BaBar in the reaction
$$
e^-e^+\to \phi(2170) \to \phi f_0(980) \gamma \ .
$$
Jefferson Lab will be able to produce it via $\gamma p\to \phi(2170) p$. 
Since this resonance has many open decay channels  ($K^+ K^-$, $K^+ K^-\pi \pi$, etc) measuring the momentum of the two charged kaons may be useful to gain insight regarding   the momentum distributions of the strange quarks in $\phi(2170)$.
This would permit testing  the presumed $3S$ quark model assignment (two Sturm-Liouville zeroes), or even documenting possible exotic  components.
Although the strange quark mass, $m_s$, is now comparable to the strong interaction strength, $\Lambda_{\rm QCD}$,  the seperation scale might still be sufficient for qualitative insight  since the strange quark momentum, $p =m_s v/\sqrt{1-v^2/c^2} $, is large according to  the relativistic virial theorem.
\\

\begin{center}
\center{\Large\em{5.  Summary}}
\\
\end{center}

As discussed in the previous pages, the Coulomb gauge formulation of QCD has
several advantages and provides an opportunity for dynamical insight regarding hadron structure.
While not readily amendable for form factor predictions, it provides a natural framework for spectroscopic studies for highly excited states as well as exotic configurations involving explicit gluonic degrees of freedom. The role of chiral symmetry has been clarified, both for light quarks,
where spontaneous symmetry breaking by condensates in the vacuum leads to a small mass pion,
as well as highly excited states, which are insensitive to this symmetry, yielding parity doublets.
Further, measuring the mass splitting between doublets allows determining the quark mass momentum dependence which connects the asymptotically free current mass to the 
much larger constituent quark value.  Finally, it has been demonstrated how the
 Frank-Condon principle can provide information about the quark momentum distribution in
 heavy quarkonium from decay measurements, some  to be performed with the Jefferson Lab 12 GeV upgrade.

\begin{center}
{\center{\small{\em Acknowledgments}}}
\end{center}
\indent

We thank the organizers for the invitation to discuss these issues at NAPP10 in Dubrovnik. Work supported by grants FPA 2008-00592, FIS2008-01323 plus 227431, Hadron-Physics2 (EU) and PR34-1856-BSCH, UCM-BSCH GR58/08, 910309, PR34/07-15875
and U. S. DOE grant DE-FG02-03ER41260.
JMT is a recipient of an FPU scholarship. 



%

\end{document}